\documentclass[twocolumn, amssymb, nobibnotes, aps, superscriptaddress,prx]{revtex4-1}

\usepackage[colorlinks=true,linkcolor=blue,citecolor=blue,urlcolor=blue]{hyperref}
\usepackage{graphicx}
\usepackage{gensymb}
\usepackage[normalem]{ulem}
\usepackage{color}
\usepackage{multirow}
\usepackage{amsmath}
\usepackage{amsfonts}
\usepackage{natbib}
\usepackage{bm}
\usepackage{textcomp}
\usepackage{orcidlink}

\def\qe{\textsc{Quantum ESPRESSO}\texttrademark}

\begin{document}

%
\title{Magnon-phonon interactions enhance the gap at the Dirac point \\ in the spin-wave spectra of CrI$_3$ two-dimensional magnets}
%

\author{Pietro Delugas\,\orcidlink{0000-0003-0892-7655}}
\affiliation{SISSA -- Scuola Internazionale Superiore di Studi Avanzati, Trieste, Italy, European Union}
\affiliation{Quantum ESPRESSO Foundation, Cambridge, CB24 6AZ, United Kingdom}

\author{Oscar Baseggio\,\orcidlink{0000-0002-5733-2001}}
\affiliation{SISSA -- Scuola Internazionale Superiore di Studi Avanzati, Trieste, Italy, European Union}

\author{Iurii Timrov\,\orcidlink{0000-0002-6531-9966}}
\affiliation{Theory and Simulation of Materials (THEOS) and National Centre for Computational Design and Discovery of Novel Materials (MARVEL), \'{E}cole Polytechnique F\'{e}d\'{e}rale de Lausanne, CH-1015 Lausanne, Switzerland}

\author{Stefano Baroni\,\orcidlink{0000-0002-3508-6663}}
\affiliation{SISSA -- Scuola Internazionale Superiore di Studi Avanzati, Trieste, Italy, European Union}
\affiliation{Quantum ESPRESSO Foundation, Cambridge, CB24 6AZ, United Kingdom}
\affiliation{CNR -- Istituto dell'Officina dei Materiali, SISSA, Trieste, Italy, European Union}

\author{Tommaso Gorni\,\orcidlink{0000-0002-7139-8429}}
\email [Corresponding author:\ ] {gornitom@gmail.com}
\affiliation{LPEM, ESPCI Paris, PSL Research University, CNRS, Sorbonne Universit\'e, 75005 Paris France, European Union}
\altaffiliation{Present address: CINECA National Supercomputing Center, Casalecchio di Reno, I-40033 Bologna, Italy, European Union}

\begin{abstract}
Recent neutron-diffraction experiments in honeycomb CrI$_3$ quasi-2D ferromagnets have evinced the existence of a gap at the Dirac point in their spin-wave spectra. The existence of this gap has been attributed to strong in-plane Dzyaloshinskii-Moriya or Kitaev (DM/K) interactions and suggested to set the stage for topologically protected edge states to sustain non-dissipative spin transport. 
We perform state-of-the-art simulations of the spin-wave spectra in monolayer CrI$_3$, based on time-dependent density-functional perturbation theory (TDDFpT) and fully accounting for spin-orbit couplings (SOC) from which DM/K interactions ultimately stem. 
While our results are in qualitative agreement with experiments, the computed TDDFpT magnon gap at the Dirac point is found to be 0.47~meV, roughly 6 times smaller than the most recent experimental estimates, so questioning that intralayer anisotropies alone can explain the observed gap. Lattice-dynamical calculations, performed within density-functional perturbation theory (DFpT), indicate that a substantial degeneracy and a strong coupling between vibrational and magnetic excitations exist in this system, providing a possible additional gap-opening mechanism in the spin-wave spectra. In order to pursue this path, we introduce an interacting magnon-phonon Hamiltonian featuring a linear coupling between lattice and spin fluctuations, enabled by the magnetic anisotropy induced by SOC. Upon determination of the relevant interaction constants by DFpT and supercell calculations, this model allows us to propose magnon-phonon interactions as an important microscopic mechanism responsible for the enhancement of the gap in the range of $\approx 4$~meV around the Dirac point of the CrI$_3$ monolayer.
\end{abstract}

\date{\today}

\maketitle

\section{Introduction}\label{sec:intro}

The van der Waals crystal CrI$_3$ was the first compound reported to display long-range magnetic order down to the monolayer limit, where it behaves as a ferromagnetic semiconductor with a Curie temperature of $45$\,K and a sizeable out-of-plane anisotropy~\cite{Huang:2017}. The impressive variety of its magnetic response to the most diverse probes has stimulated an intense research effort aiming to better understand and exploit its low-dimensional magnetism. Indeed, spintronics applications are envisaged for bilayer-CrI$_3$, a layered antiferromagnet which can be gradually turned ferromagnetic by applied pressure~\cite{Li:2019,Song:2019}, electrical field~\cite{Huang:2018,Jiang:2018a}, or electrostatic doping~\cite{Jiang:2018b}, thus realizing an intrinsic spin filter tunnel junction with a reported high figure of merit~\cite{Wang:2018,Song:2019,Klein:2018,Kim:2018}. Moreover, the optical properties of ultrathin CrI$_3$ have been shown to be exceptionally sensitive to its magnetic state~\cite{Seyler:2018,Sun:2019,Huang:2020}, stimulating new ideas for optoelectronic~\cite{Jiang:2018c} and photovoltaic~\cite{Zhang:2019} devices. Furthermore, the insulating character of this material strongly reduces Landau damping, thus ensuring a magnon lifetime longer than that of any known 2D metal by one order of magnitude~\cite{Cenker:2020}, with tremendous implications in the field of magnonics~\cite{Zakeri:2020}. Even more interestingly, the honeycomb structure of the 2D Cr lattice is such that a system of Heisenberg magnets localized at the atomic sites would feature two spin-wave dispersion crossing at the $K/K'$ corners of the Brillouin zone (BZ). Anisotropic exchange interactions --such as those caused by Spin Orbit Coupling (SOC)-- may open a gap at the Dirac points by breaking the inversion symmetry and developing topologically protected edge spin waves able to sustain dissipation-less spin transport~\cite{Fransson:2016, Kim:2016, Owerre:2016, Pershoguba:2018}, in full analogy with fermionic topological states originally proposed to occur in graphene~\cite{Kane2005}.

Indeed, inelastic neutron scattering (INS) has provided evidence of a gap of a few meV at the Dirac points in the spin-wave spectrum of quasi-2D CrI$_3$ samples~\cite{Chen:2018,Chen:2021}. While the existence of multiple magnon states at the center of the BZ has been confirmed by (magneto-) Raman spectroscopy \cite{Jin:2018,McCreary:2020,Cenker:2020} the nature and very existence of a gap at the zone border is still controversial. 
Early suggestions explained the occurrence of this gap in terms of SOC-enabled inversion-symmetry-breaking exchange interactions, such as second-nearest-neighbor Dzyaloshinskii-Moriya (DM)~\cite{Chen:2018,Chen:2021} or first-nearest-neighbor Kitaev (K)~\cite{Lee:2020}. While this scenario has received support from itinerant-electron models based on tight-binding Hamiltonians~\cite{Costa:2020}, the robustness of these conclusions is as strong as the reliability of the semiempirical SOC and screened Coulomb repulsion parameters on which they are based. 
Moreover, the direct estimation from first-principles of the anisotropic exchange couplings has yielded quite diverse results according to the details of the downfolding methodology~\cite{Soriano:2020}, with the most recent calculations pointing towards very weak~\cite{Kvashnin:2020}, if not negligible~\cite{Pizzochero:2020}, DM/K interactions, thus further questioning their role in the opening of the observed gap.
Recently, \emph{ab initio} calculations of magnetic excitations in CrI$_3$ have confirmed the weak nature of SOC-induced anisotropies in monolayer CrI$_3$~\cite{Olsen:2021} and underlined the relevance of interlayer interactions in bulk CrI$_3$~\cite{Ke:2021}.
The authors of the present work have recently shown by means of relativistic calculations based on time-dependent density-functional perturbation theory (TDDFpT) that, while SOC is the primary cause of the gap, its combination with inter-layer interactions may more than double its actual width, so accounting for more than 50\% of its experimental estimate~\cite{Gorni:2023}.
In order to fathom the nature of this discrepancy, it is crucial to assess whether other microscopic mechanisms might be at play in the gap opening.

In this work we provide evidence that strong spin-lattice interactions are present in monolayer CrI$_3$ and may manifest in a polariton-like hybridization between magnetic and vibrational modes \emph{near}, but not quite \emph{at}, the Dirac points, which may enhance the opening of the gap.
Our first step is the evaluation of the INS magnetic cross section of monolayer CrI$_3$, performed at clamped nuclei by means of full-fledged TDDFpT and wholly accounting for relativistic effects~\cite{Walker:2006,*Rocca:2008,*Timrov:2013,Gorni:2018}. Our approach, not relying on any adiabatic spin decoupling \cite{Gebauer2000}, avoids the intricacies of downfolding to an effective spin model and so includes the complexity of SOC-induced exchange couplings directly into the excitation spectra without the need of introducing any semi-empirical parameters. Our results reveal two dispersive magnon branches with quite different cross-section intensities, in fair agreement with INS data~\cite{Chen:2018,Chen:2021}, except for too large a predicted band width, which is a consequence of a common shortcoming of the local spin-density (LSD) and related approximations, which are known to predict too large a spin stiffness~\cite{Yin:2011,Singh:2019}.

We clearly resolve a sizeable Goldstone gap at the zone center ($\approx 1.3$~meV), the hallmark of magnetic anisotropy, and a smaller gap at the Dirac points of about $\approx 0.47$~meV, corroborating previous estimates of very weak DM/K interactions in monolayer CrI$_3$~\cite{Pizzochero:2020, Ke:2021}. 
Next, following recent indications that a strong spin-lattice coupling may exist in ultrathin CrI$_3$~\cite{McCreary:2020,Webster:2018,Rodriguez-Vega:2020}, we determine the phonon dispersions in this system, using (static) density-functional perturbation theory (DFpT)~\cite{Baroni:1987a,*Giannozzi:1991a,*Baroni:2001}, and reveal that a bundle of flat phonon bands intersects the acoustic magnon band just below the experimentally observed energy of the Dirac magnon.
Building on this finding we identify, out of the phonon bundle, a flat branch with frequency $\hbar\omega\approx 14~\text{meV}$ displaying the strongest spin-lattice coupling, in fair agreement with the position of the observed gap at $\hbar\omega\approx 12$\,meV in CrI$_3$ thin crystals~\cite{Chen:2018,Chen:2021}.
This coupling, which is enabled by SOC, can be captured by a generalized Heisenberg Hamiltonian whose eigenstates are spin-lattice polaritons featuring a gap where the free magnon and phonon bands cross. 
We show that magnon-phonon interactions open a gap of about $\Delta \approx 1.5$\,meV at $q\approx 0.6 q_K$ along the $\Gamma K$ line. The magnitude of the gap depends on the strength of the spin-lattice coupling, whereas the shape and location of the locus of points in the BZ where it occurs depends on the details of the phonon and magnon dispersions. 
While thirty years of successful practice of DFpT give us confidence in the accuracy of our predictions for the phonon bands, density-functional theory, particularly when based on LSDA or spin-polarized generalized-gradient approximation (GGA), is known to overestimate the magnitude of magnetic moments and exchange interactions, as recently reported also for bulk CrI$_3$~\cite{Ke:2021,Gorni:2023}.
In this very case, it has been shown that this shortcoming can be significantly redressed by on-site Hubbard corrections for Cr $3d$ states~\cite{Ke:2021,Olsen:2019}, which result in a quite rigid renormalization of the magnon bandwidth.
Because of this, the exchange interactions that define our Heisenberg spin-lattice Hamiltonian have been renormalized to the experimental spin stiffness, as discussed in Appendix~\ref{App:Jexch}.

This paper is organized as follows. Section~\ref{sec:tddfpt} summarizes our theoretical framework and presents the results of our TDDFpT calculations; in Sec.~\ref{sec:phonons} our DFpT calculations of the phonon dispersion and supercell calculations of the magnon-phonon coupling are reported; Sec.~\ref{sec:magnon-phonon} discusses the implications of the predicted spin-lattice polaritons on the magnon dispersion based on a generalized Heisenberg Hamiltonian. Finally, Sec.~\ref{sec:Conclusions} presents our conclusive remarks.

\section{Inelastic neutron-scattering cross sections from TDDFpT}
\label{sec:tddfpt}

In INS experiments a neutron beam with wavevector $\mathbf{k}_i$ and energy $E_i$ is inelastically scattered by the target sample to a final state characterized by the wavevector $\mathbf{k}_f = \mathbf{k}_i - \bm{q}$ and energy $E_f = E_i - \hbar \omega$, where $\hbar\bm{q}$ and $\hbar \omega$ are the momentum and energy transferred to the sample, respectively. In the first Born approximation~\cite{Halpern:1938aa,Blume:1963aa}, the double-differential cross section corresponding to magnetic excitations of electrons can be written in the compact form as \cite{Placzek:1954,VanHove:1954}:
\begin{equation}
    \frac{d^2 \sigma}{d\Omega d\omega} = \frac{\hbar}{\pi}\left(\frac{g_n e}{2\hbar}\right)^2\frac{k_f}{k_i} \, S(\bm{q},\mathbf{\omega}) \,, \label{eq:cross_section}
\end{equation}
where
\begin{equation}
    S(\bm{q},\mathbf{\omega}) = -{\rm Im}\,{\rm Tr} \bigg[ {\boldsymbol P}^{\perp}(\bm{q}) \, {\boldsymbol \chi}(\bm{q},\bm{q}; \omega) \bigg] \,. \label{eq:S_def}
\end{equation}
Here, $-e$ and $g_n \approx 3.826$ are the electron charge and the neutron $g$-factor, respectively, ${\boldsymbol P}^{\perp}(\bm{q})$ is the $3 \times 3$ matrix, $P_{\alpha\beta}^{\perp}(\bm{q}) = \delta_{\alpha\beta} - q_{\alpha}q_{\beta}/q^2$ (with $\alpha , \beta = x,y,z$), which is a projector onto the plane perpendicular to the direction of $\bm{q}$, and $\bm{\chi}(\bm{q},\bm{q};\omega)$ is the $3 \times 3$ spin susceptibility matrix. The poles of $S(\bm{q},\mathbf{\omega})$ are fingerprints of the spin excitations of the system, both of the magnon and Stoner type, thus allowing one to characterize various magnetic spectroscopies, either in the bulk, as probed by INS, or at surfaces, as probed by spin-polarized electron energy loss spectroscopy~\cite{Gokhale:1992}. A fully \emph{ab~initio} determination of $S(\bm{q},\mathbf{\omega})$ requires the computation of the dynamical spin susceptibility by using, \emph{e.g.}, TDDFpT~\cite{Savrasov:1998, Lounis:2011, Buczek:2011b, Rousseau:2012, dosSantosDias:2015, Gorni:2016, Wysocki:2017, Cao:2018, Gorni:2018}.

We have computed the INS magnetic spectrum of monolayer CrI$_3$ using the Liouville-Lanczos approach to TDDFpT of Ref.~\cite{Gorni:2018}, as implemented in the \texttt{turboMagnon} code~\cite{Gorni:2022b} within the LSDA in the plane-wave pseudopotential framework of the \qe\ suite of codes~\cite{Giannozzi:2009,*Giannozzi:2017,*Giannozzi:2020}. We stress that our implementation applies to a general non-collinear spin-polarized framework, essential in materials containing heavy elements, such as CrI$_3$, where SOC is expected to play an important role~\cite{Soriano:2020}. Further technical details of our computations are presented in Appendix~\ref{App:computational}, 

\begin{figure}[t]
    \begin{center}
        \includegraphics[width=0.48\textwidth]{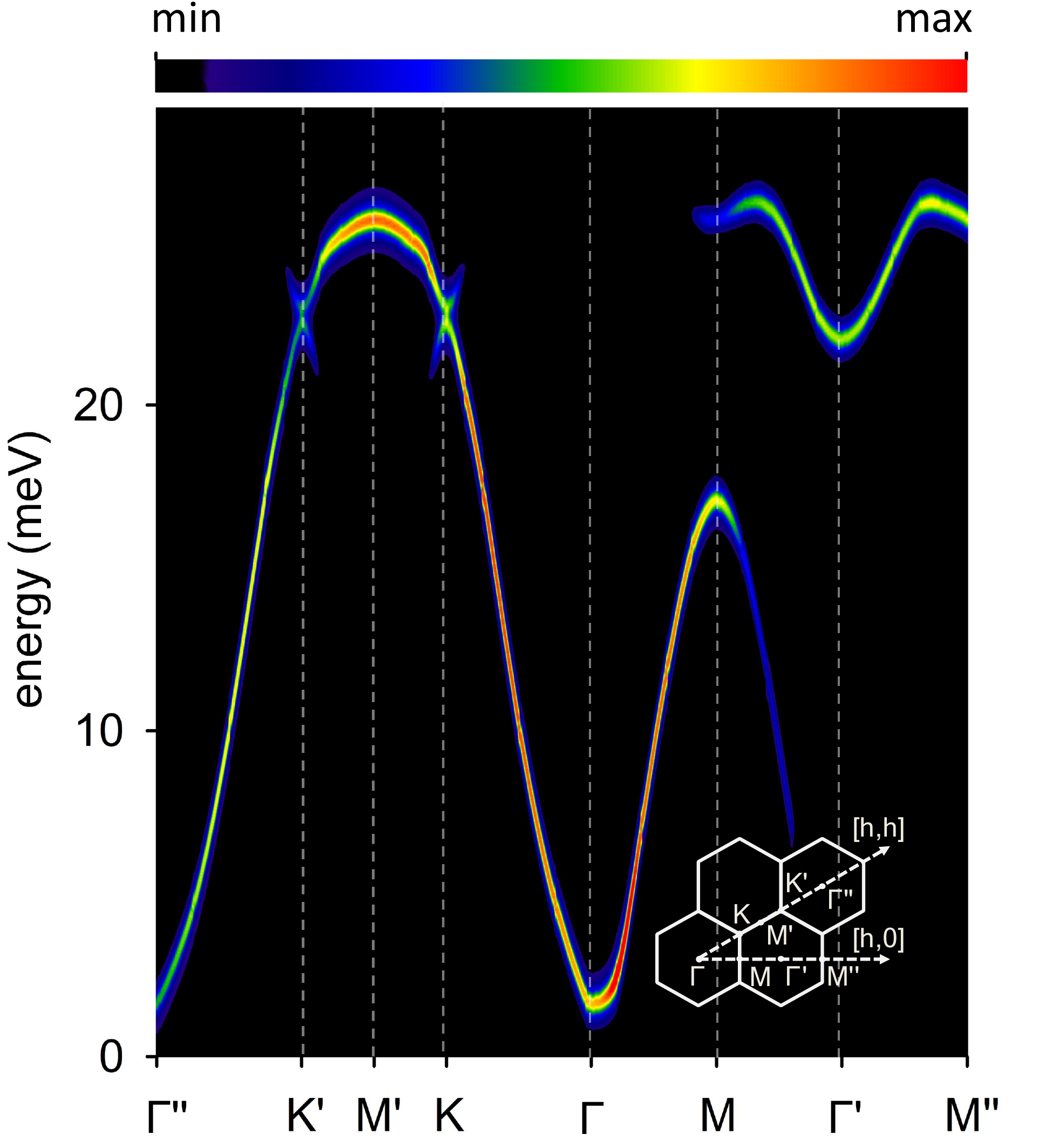}
        \caption{Magnon dispersions along high-symmetry directions in the BZs computed using TDDFpT including SOC in a ferromagnetic CrI$_3$ monolayer. The color code describes the intensities of peaks in the magnetic spectrum. The inset shows adjacent BZs and the high-symmetry directions in the $(h,k)$ plane and labels of high-symmetry points.
        }
        \label{fig:TDDFpT_spectrum}
    \end{center}
\end{figure}

The spin-wave spectrum of monolayer CrI$_3$ thus obtained is reported in Fig.~\ref{fig:TDDFpT_spectrum}: two magnon branches, which can be interpreted as arising from two Heisenberg moments arranged in a honeycomb lattice, can be clearly identified. In particular, an intense acoustic band is present in the first BZ along the $\Gamma M$ direction, whereas the optical branch takes over when entering the second BZ. Our results can be summarized as follows:
\emph{i)}~we overestimated the spin stiffness of the system, resulting in too broad a band width of $\approx 28$ meV (consistent with the linear-response DFT-based calculations of Ref.~\cite{Ke:2021}, reporting a value of $\approx 31$~meV in bulk CrI$_3$ without SOC), to be compared with an experimental value for a thin CrI$_3$ crystal of $\approx 20$~meV~\cite{Chen:2018,Chen:2020,Chen:2021};
\emph{ii)}~a Goldstone gap of 1.3~meV is clearly detected at the BZ center;
\emph{iii)}~a small gap ($\approx 0.5$ meV) is also detected at the Dirac point ($K$).
While too large a spin stiffness is a common feature of LSDA- and GGA-based models~\cite{Yin:2011,Singh:2019,Ke:2021}, we stress that our theoretical framework naturally accounts for the SOC resulting in the observed gaps at the $\Gamma$ and $K$ points. 
We conclude that SOC-induced DM/K interactions at clamped nuclei, which are implicitly accounted for in our \emph{ab initio} approach, are likely too small to account for the origin of the observed gap at the Dirac ($K$) point alone, as it was already suggested in previous theoretical studies~\cite{Kvashnin:2020,Pizzochero:2020,Gorni:2023}.
In a recent work of ours \cite{Gorni:2023} it is pointed out that inter-layer interactions may account for much of the gap observed in multi-layer samples. In the following we  give evidence that strong spin-lattice couplings may provide an additional mechanism to enhance the magnitude of the gap at the Dirac point in CrI$_3$.

\section{PHONON DISPERSIONS AND SPIN-LATTICE COUPLING}
\label{sec:phonons}

\begin{figure}[t]
    \centering
    \includegraphics[width=0.48\textwidth]{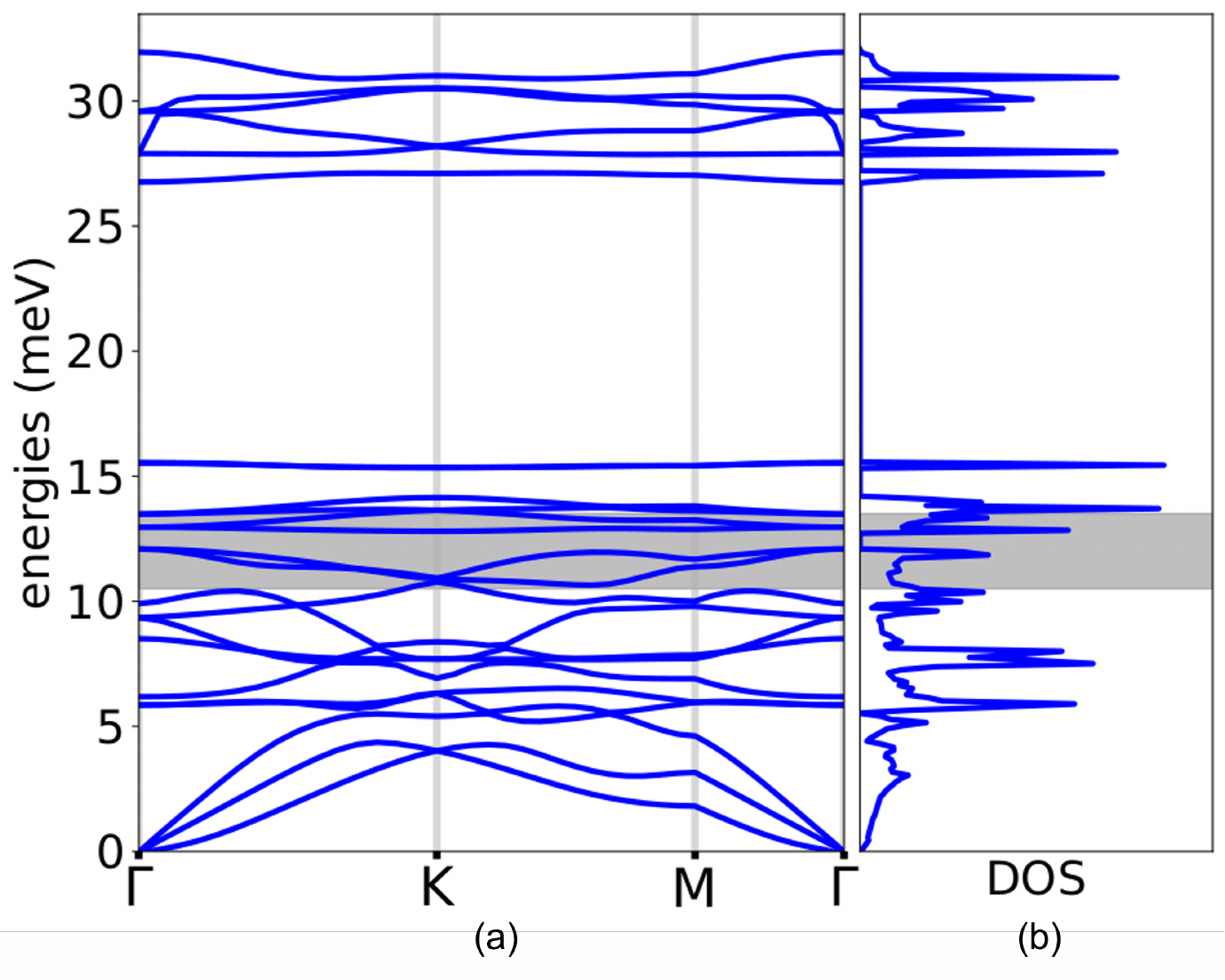}
    \caption{(a)~Phonon dispersions computed using DFpT; the shaded area highlights the experimental inter-band magnon gap~\cite{Chen:2021}; we show in Fig.~\ref{fig:couplings} the spin-lattice couplings for the modes that build up the strong peak of the vibrational density of states in this region; (b)~vDOS corresponding to the phonon dispersion reported in panel (a). }
    \label{fig:phonons}
\end{figure}

\begin{figure*}[t]
    \centering
    \includegraphics[width=0.95\textwidth]{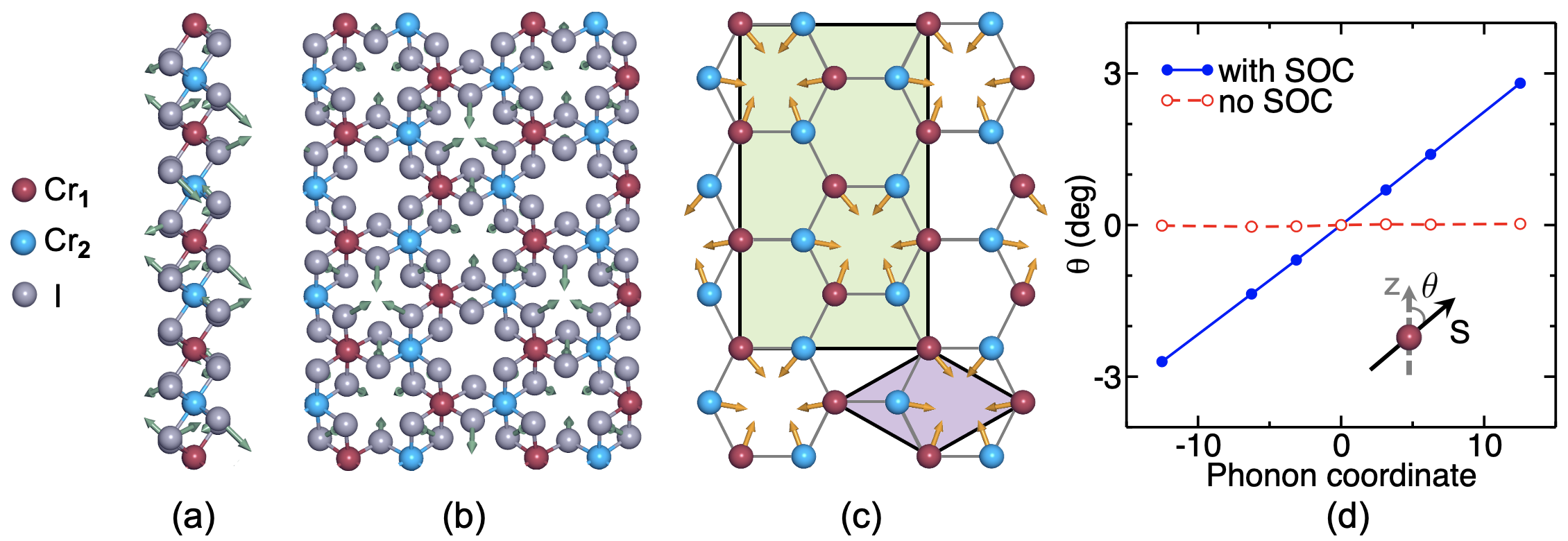}
    \caption{
        Side (a) and top (b-c) views of the CrI$_3$ monolayer. Cr atoms belonging to different sublattices are depicted with different colors (blue and red). In panels (a-b) the green arrows indicate the displacement pattern of the 16-th vibrational mode at the $K$ point of the BZ, which is the one with the strongest coupling with the atomic magnetic moments. As a consequence of this coupling, the atomic moments acquire an in-plane component whose magnitude is proportional to the lattice distortion, as depicted by the yellow arrows in panel (c). Panel (d) displays the dependence of the resulting tilt angle on the phonon amplitude, $\xi^{16}(\bm{q}_K)$ [see Eq.~\eqref{eq:ph-normal-modes}, in units $\mathrm{\AA\times\sqrt{\mathrm{AMU}}}$], accounting for or neglecting SOC. The light-purple and light-green shaded areas indicate the unit cells of the undistorted and distorted lattices, respectively.}
    \label{fig:mode-16-comparison}
\end{figure*}

It was recently pointed out that CrI$_3$ features both a strong vibrational density of states (vDOS) at $\approx 14$\,meV---in close proximity to the position of the magnon gap observed at $\approx 12.5$~meV in the CrI$_3$ thin crystals---and a strong spin-lattice coupling~\cite{Webster:2018}. The occurrence of these two facts has led to the speculation that hybrid magneto-vibrational excitations~\cite{Kittel:1949,Kittel:1958} may occur in this system~\cite{Kvashnin:2020}, resulting in a sort of polaritonic mixing between magnon and phonon bands.

In order to elaborate on this surmise, we have started by computing the phonon spectrum of CrI$_3$. For further reference, and in order to fix the notation, let us first state the lattice Hamiltonian in the harmonic approximation:
\begin{align}
    H_{\rm ph} = \sum_{ns} \frac{\bm{p}_{ns}^{2}}{2M_s} + \frac{1}{2}\sum_{ns \ne mt} \bm{u}_{ns} \cdot \bm{\mathcal{K}}_{st}(\bm{R}_{nm}) \cdot \bm{u}_{mt},
    \label{eq:ph-Ham}
\end{align}
where $n$ and $m$ enumerate the elementary cells of the crystal lattice, $\bm{R}_{nm}=\bm{R}_{n}-\bm{R}_{m}$ is the distance between two such cells, $\bm{u}_{ns}$ is the displacement of the $s$-th atom in the $n$-th unit cell, $\bm{p}_{ns}$ and $M_s$ are its momentum and mass, respectively, and $\bm{\mathcal{K}}_{st} (\bm{R}_{nm})$ is the matrix of the interatomic force constants, which we compute using DFpT~\cite{Baroni:2001}. Bold symbols indicate 3-vectors (Cartesian indices are suppressed and a dot ``$\cdot$'' stands for a scalar product), whereas bold calligraphic symbols are $3\times 3$ tensors. The vibrational normal modes are obtained from the eigenvalue equation (a tilde, ``$\tilde{\phantom{X}}$'', on top of various quantities indicates their Fourier transforms, when needed):
\begin{align}
    \sum_{t} \bm{\widetilde{\mathcal{K}}}_{st}(\bm{q}) \cdot \bm{e}_{t}^\mu(\bm{q}) = M_s\Omega^2_\mu(\bm{q}) \bm{e}_{s}^\mu(\bm{q}), \label{eq:vibNormMod}
\end{align}
where $\bm{q}$ is the phonon wavevector, $\Omega_\mu(\bm{q})$ and $\bm{e}_{s}^\mu(\bm{q})$ are the eigenvalues (frequencies) and eigenvectors of the $\mu$-th vibrational mode.
The atomic displacements in the $\mu$-th normal mode are defined in terms of the normalized eigenvectors of Eq.~\eqref{eq:vibNormMod} $\bigl (\sum_s M_s  \bm{e}_{s}^{\mu *}(\bm{q}) \cdot \bm{e}^\nu_{s}(\bm{q}) =\delta_{\mu\nu} \bigr ) $ as:
\begin{align}
    \bm{u}_{ns} = \frac{1}{N}\sum_{\bm{q}\mu} \mathrm{e}^{i\bm{q}\cdot\bm{R}_n}\xi^\mu(\bm{q}) \bm{e}^\mu_{s}(\bm{q}),
    \label{eq:ph-normal-modes}
\end{align}
where $\xi^\mu(\bm{q})=\sum_{ns} M_s \mathrm{e}^{-i\bm{q}\cdot\bm{R}_n} {\bm{u}}_{ns} \cdot \bm{e}^{\mu *}_{s}(\bm{q}) $ is the amplitude of the $\mu$-th normal mode, and $N$ the number of unit cells in the crystal. For future reference, we remark that the mode length $\xi^{\mu}(\bm{q})$  incorporates the reduced mass of the specific mode, and its dimensions are a length times the square root of a mass (say, $\mathrm{\AA\times\sqrt{\mathrm{AMU}}}$). We solved the eigenproblem of Eq.~\eqref{eq:vibNormMod} within DFpT using LSDA, including SOC self-consistently. All the relevant technical details of the computations reported in this section are presented in Appendix~\ref{App:computational}.

In Fig.~\ref{fig:phonons} we display our computed phonon dispersions and vDOS for monolayer CrI$_3$.
Our results confirm that a high vDOS exists in the energy range where a magnetic gap is observed, thus pointing at magnon-phonon interactions as a candidate mechanism that could affect the magnon gap around the Dirac point. 
In order to ascertain whether this can indeed be the case, we have computed the dependence of the crystal magnetization on the lattice distortion along the normal modes in the relevant energy range.
We find that the vDOS peak is populated with vibrational modes strongly coupled with the magnetization, as reported in Fig.~\ref{fig:Theta-of-Q} of Appendix~\ref{App:spin-lattice-constants}.
The most intense spin-lattice coupling is found with the 16-th normal mode, whose energy is $\hbar\Omega_{16}(\bm{q}_K) \approx 14$\,meV. Displacing the atoms along its eigenvectors $\bm{u}_{ns}$ at $\bm{q}=\bm{q}_K$, shown in green in Figs.~\ref{fig:mode-16-comparison}(a) and \ref{fig:mode-16-comparison}(b), induces a tilt of the crystal magnetization with respect to the easy-magnetization ($z$) axis according to the pattern presented in Fig.~\ref{fig:mode-16-comparison}(c). In Fig.~\ref{fig:mode-16-comparison}(d) we report the magnitude of such a  tilt angle as a function of the amplitude of the lattice distortion. 
Not unexpectedly, no such dependence is detected when neglecting SOC, which is the origin of the magnetic anisotropy. When fully accounting for SOC, instead, a strong linear spin-lattice coupling is observed, further confirming that SOC-mediated spin-lattice couplings may play a relevant role in the magnon dispersion around the $K$ point.

\section{Mixed spin-lattice excitations}
\label{sec:magnon-phonon}

\subsection{Spin-lattice Hamiltonian}
\label{sec:SL_Hamiltonian}

In order to derive the minimal Hamiltonian accounting for a spin-lattice coupling, we consider the most general quadratic spin Hamiltonian, whose interaction parameters depend on the spin-spin distance:
\begin{multline}
    H_{\rm sp} = -\frac{1}{2}\sideset{}{'}\sum_{ns \neq mt} \bm{S}_{ns} \cdot \bm{\mathcal{J}}_{st}(\bm{R}_{nm}) \cdot \bm{S}_{mt} \\ -\sideset{}{'}\sum_{ns} \bm{S}_{ns} \cdot \bm{\mathcal{D}}_{s} \cdot \bm{S}_{ns} \, ,
    \label{eq:spin-ham}
\end{multline}
where $\bm{S}_{ns}$ is a classical spin residing at the $sn$ lattice site, $\bm{\mathcal{J}}_{st}(\bm{R}_{nm})$ the exchange couplings and $\bm{\mathcal{D}}_{s}$ the onsite magnetic anisotropy. The primed summations run on the magnetic sites only, in contrast to the summations in Eq.~\eqref{eq:ph-Ham}, which run over all the atomic positions.
Both the exchange couplings and the onsite magnetic anisotropy depend implicitly on the atomic displacements and can be expanded in powers of $\bm{u}_{ns}$.
In the undistorted geometry ($\bm{u}_{ns}=0$),
magnetic interactions are modeled
only via isotropic exchange $J_{st}(\bm{R}_{nm}) = {\rm Tr}\Big[\bm{\mathcal{J}}_{st}(\bm{R}_{nm})\Big]/3$ and the onsite anisotropy $\mathcal{D}^{zz}_s$ couplings whose values are derived by means of supercell calculations, as reported in Appendix~\ref{App:Jexch}.
In view of the weakness of the inter-site anisotropies responsible for the $0.47$\,meV gap found in our TDDFpT calculations, we chose not to include them in our lattice model and focus on the effect of magnon-phonon couplings only. 
Considering the smallness of these two effects, we expect that, if considered together, they would add linearly.

At  zeroth order in the spin-lattice interactions, the vibrational and magnetic normal modes are decoupled, the latter being solutions of the secular equation:
\begin{equation}
    S\sideset{}{'}\sum_t \bigg[ \delta_{st} I_s - \tilde{J}_{st}(\bm{q}) \bigg] f_t^{\nu}(\bm{q}) = \omega_{\nu}(\bm{q}) f_s^{\nu}(\bm{q}) , \label{eq:spin-normal-modes}
\end{equation}
where $I_s = 2D^{zz}_s + \sum'_{mt} J_{st}(\bm{R}_{0m})$, and $S = |\bm{S}_{ns}|$. 
The eigenvectors are normalized according to  $\sum_s f^{\mu*}_{s}(\bm{q}) f^{\nu}_{s}(\bm{q}) = S\delta_{\mu\nu}$ and are directly related to the spin components in the $xy$ plane via
\begin{equation}
    S^+_{ns} = \frac{1}{N} \sum_{\bm{q}\nu} \mathrm{e}^{i\bm{q}\cdot\bm{R}_n} \, \eta^{\nu}(\bm{q}) f^{\nu}_s(\bm{q}) \, , \label{eq:mag-normal-modes}
\end{equation}
with $S_{ns}^{\pm} = S_{ns}^x \pm i S_{ns}^y$, and $\eta^{\nu}(\bm{q})$ is the amplitude of the $\nu$-th normal mode. 

To lowest-order in $\bm{u}_{ns}$, the corrections to the Hamiltonian in Eq.~\eqref{eq:spin-ham} result in a linear coupling between the spin and lattice degrees of freedom (see Appendix~\ref{App:Hamiltonian_mp} for a derivation), reading:
\begin{equation}
    H_{\rm sl} = -\sum_{\bm{q}\mu\nu} \lambda_{\mu\nu}(\bm{q}) \sqrt{\omega_{\nu}(\bm{q})} \xi^{\mu}(\bm{q}) \eta^{\nu *}(\bm{q}) + \mathrm{c.c.} \, , \label{eq:spin-lattice-ham}
\end{equation}
where the factor $\sqrt{\omega_{\nu}(\bm{q})}$ has been introduced so that $\lambda_{\mu\nu}(\bm{q})$ has the dimension of a frequency. 

The exchange interactions in Eq.~\eqref{eq:spin-normal-modes}, $\tilde J_{st}(\bm{q})$,  as well as the $\lambda_{\mu\nu}(\bm{q})$ coupling constants in Eq. \eqref{eq:spin-lattice-ham} corresponding to different phonon-magnon pairs, $(\mu,\nu)$, can be estimated using
standard clamped-ion DFT calculations along a distortion pattern, as explained in Appendix~\ref{App:spin-lattice-constants}. In Fig. \ref{fig:couplings} we report the magnitude of the couplings between the acoustic magnon and the phonons in the energy range from 8 to 15~meV, thus estimated at the $K$ point in the BZ, showing that a particularly strong coupling exists with the 16-th phonon mode
at $\approx 14$ meV, very close to the position of the observed gap. In the next section we show that such a magnon-phonon coupling may significantly affect the magnon dispersion around the $K$ point.

\begin{figure}[t]
    \begin{center}
        \includegraphics[width=0.48\textwidth, trim = 5 0 5 0 , clip]{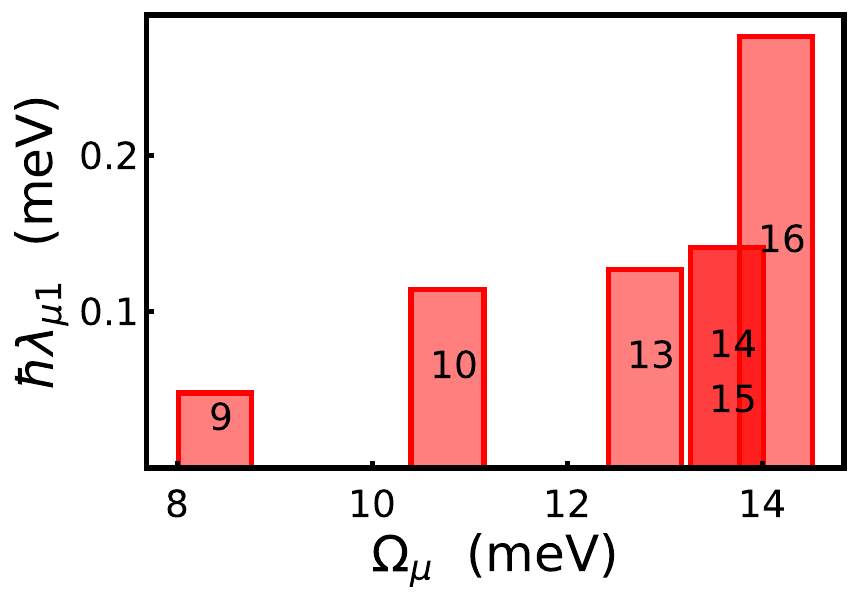}
        \caption{Absolute value of the magnon-phonon coupling constants, $\lambda_{\mu 1}(\bf{q})$, for different phonon modes of frequency $\Omega_\mu$ ($\mu=9,..., 16$) and for the acoustic magnon ($\nu=1$), computed at the Dirac point, $\bm{q}=\bm{q}_K$, in monolayer CrI$_3$. The $11-{\rm th}$ and $12-{\rm th}$ phonon modes do not display any linear coupling with magnons, as reported in Fig.~\ref{fig:Theta-of-Q} and discussed in App.~\ref{App:spin-lattice-constants}. }
        \label{fig:couplings}
    \end{center}
\end{figure}

\begin{figure}[t]
    \begin{center}
        \includegraphics[width=0.48\textwidth]{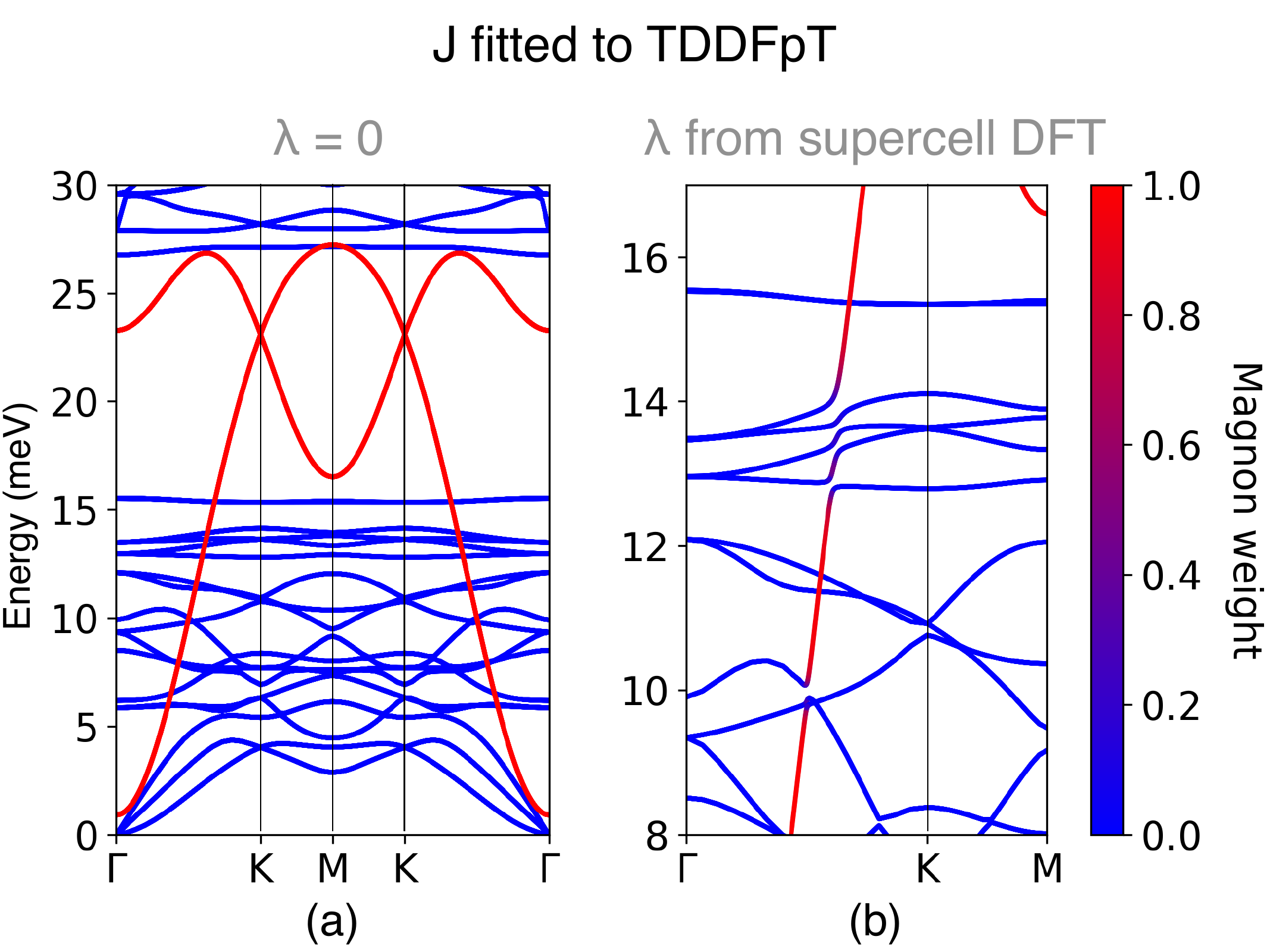}
        \includegraphics[width=0.48\textwidth]{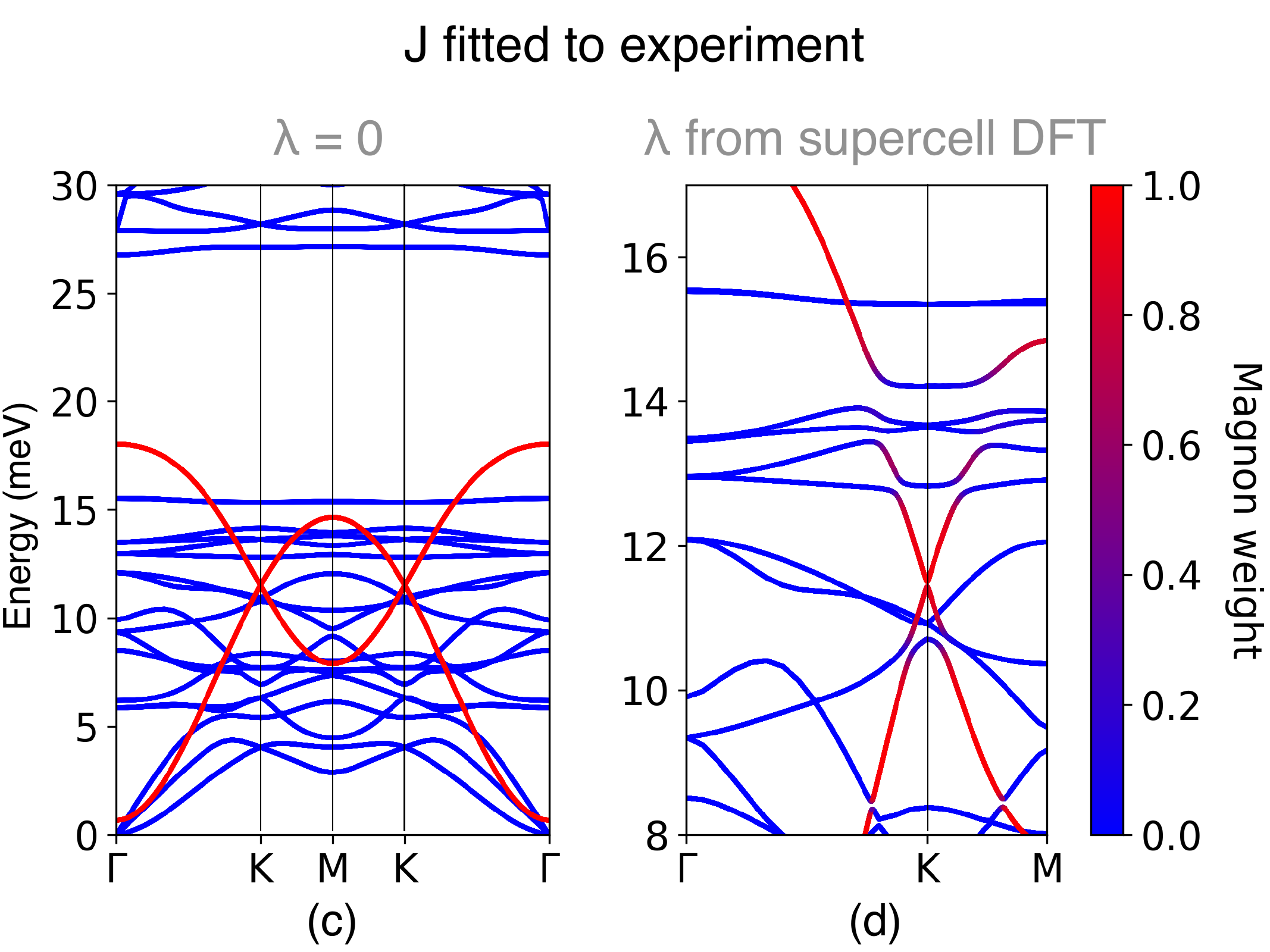}
        \caption{
        Theoretical magnon (red) and phonon (blue) dispersions in monolayer CrI$_3$, as obtained by diagonalizing the lattice Hamiltonian, Eq. \eqref{eq:quantum-ham}. In panels (a) and (c) the spin-lattice coupling has been disregarded, while it is accounted for in panels (b) and (d). The magnetic exchange parameters in Eq. \eqref{eq:quantum-ham} have been fitted to our TDDFpT calculations in panels (a) and (b) and to experiments \cite{Chen:2018} in panels (c) and (d).
        } \label{fig:polaritons-theo}
    \end{center}
\end{figure}

\subsection{Spin-lattice polaritons}

The normal modes of Eq.~\eqref{eq:spin-normal-modes} yield the frequencies and polarization patterns of the free spin waves, \emph{i.e.} of the magnetic excitations resulting from the neglect of the spin-lattice interactions embodied in Eq.~\eqref{eq:spin-lattice-ham}.
In order to quantify the impact of these interactions on the spin-wave dispersion, we quantize the spin-lattice Hamiltonian and obtain:
\begin{multline}
        \hat{H} =\: 
        \sum_{\mu\bm{q}} \hbar\Omega_{\mu}(\bm{q}) \bigg( \hat{a}^{\dag}_{\mu\bm{q}}\hat{a}_{\mu\bm{q}} + \frac{1}{2} \bigg) \\
         + \sum_{\nu\bm{q}} \hbar\omega_{\nu}(\bm{q}) \bigg( \hat{b}^{\dag}_{\nu\bm{q}}\hat{b}_{\nu\bm{q}} + \frac{1}{2} \bigg) \\
         - \sum_{\mu\nu\bm{q}} \bigg[ \hbar\bar{\lambda}_{\mu\nu}(\bm{q}) \bigg( \hat{a}_{\mu\bm{q}} \hat{b}^{\dag}_{\nu\bm{q}} + \hat{a}^{\dag}_{\mu-\bm{q}} \hat{b}^{\dag}_{\nu\bm{q}} \bigg) + {\rm h.c.} \bigg] \, ,
    \label{eq:quantum-ham}
\end{multline}
where $\hat{a}_{\mu\bm{q}}$ and $\hat{b}_{\nu\bm{q}}$ are the annihilation operators of phonons and magnons, respectively, and $\bar{\lambda}_{\mu\nu}(\bm{q}) =\lambda_{\mu\nu}(\bm{q}) \sqrt{2\omega_{\nu}(\bm{q})/ \Omega_{\mu}(\bm{q})}$.
The phonon frequencies $\Omega_{\mu}(\bm{q})$ are interpolated from our DFpT calculations, whereas the magnon frequencies $\omega_{\nu}(\bm{q})$ are derived from two different spin models, as specified in the following.
The $\lambda_{\mu\nu}(\bm{q})$ couplings are estimated at the Dirac point as detailed in the previous section and considered constant throughout the BZ. In the following we neglect the terms $\hat{a}^{\dag}_{\mu-\bm{q}}\hat{b}^{\dag}_{\nu\bm{q}}$ that do not conserve the number of quasiparticles and are thus expected to yield higher-order corrections with respect to the number-conserving terms.

\begin{figure}[t]
    \begin{center}
        \includegraphics[width=0.43\textwidth]{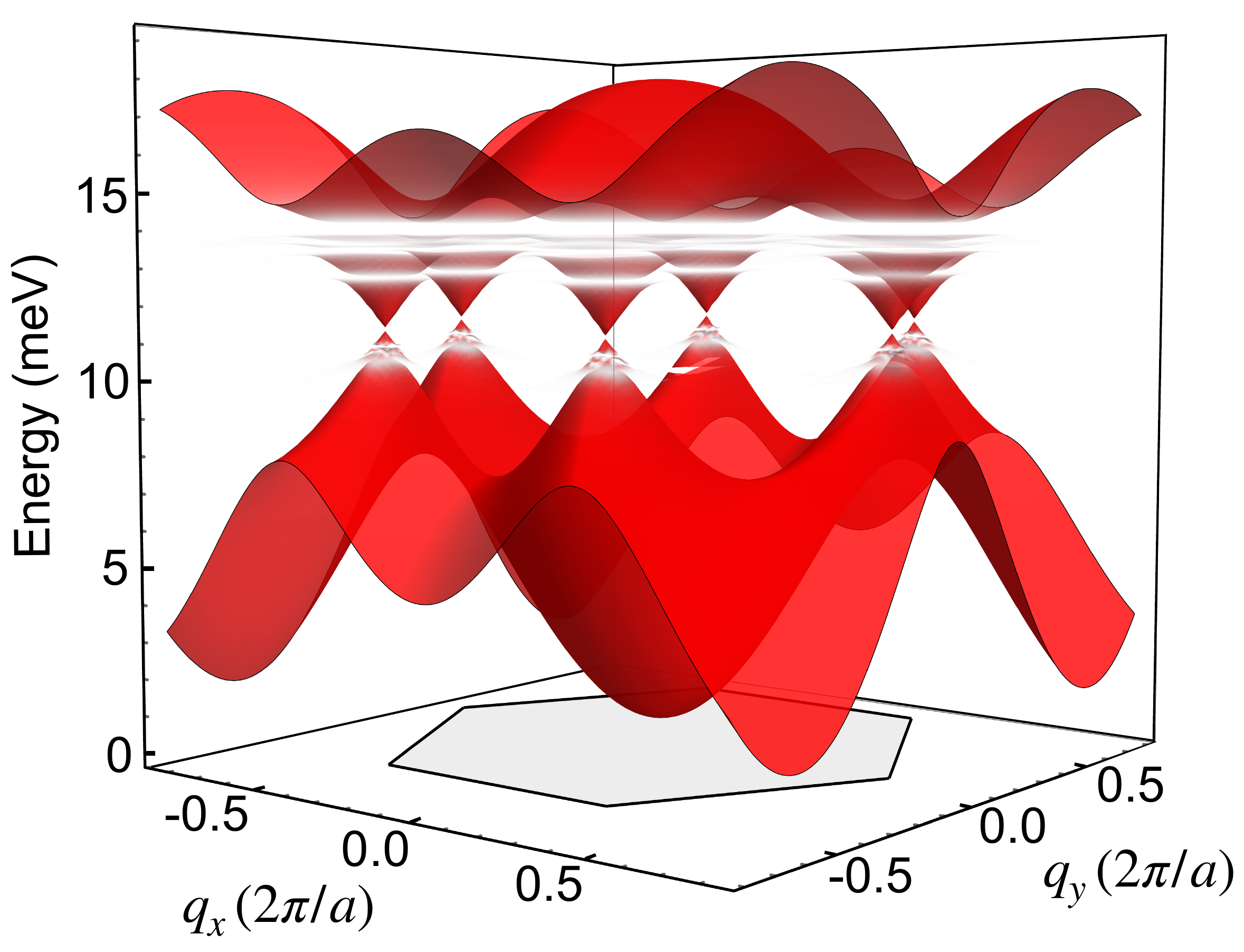}
        \caption{The 3D view of the magnon dispersion of the CrI$_3$ monolayer computed using the model Hamiltonian, Eq.~\eqref{eq:quantum-ham}, including magnon-phonon interactions. The color code goes from red (100\% magnon) to transparent (100\% phonon). The 2D BZ is shown at the bottom plane of the figure. The magnon momentum wavector is in units of $2\pi/a$, where $a$ is the lattice parameter.
        } \label{fig:3D}
    \end{center}
\end{figure}

The eigenmodes of the Hamiltonian in Eq.~\eqref{eq:quantum-ham} are mixed magnon-phonon quasiparticles. As their nature is remnant of the mixed phonon-photon or exciton-photon modes commonly known as \emph{polaritons}, we dub them \emph{spin-lattice polaritons}. In Fig.~\ref{fig:polaritons-theo} we report the energy dispersions of spin-lattice polaritons obtained by diagonalizing the Hamiltonian in Eq.~\eqref{eq:quantum-ham} in the number-conserving approximation. We consider two different sets of free-magnon frequencies, $\omega_\nu(\bm{q})$: in Figs.~\ref{fig:polaritons-theo}(a) and (b) the one obtained by using the exchange parameters fitted to our TDDFpT results, in Figs.~\ref{fig:polaritons-theo}(c) and (d) the one obtained by using the exchange parameters fitted to the INS data of Ref.~\cite{Chen:2018}.
We remark that these magnon frequencies were obtained by neglecting the DM/Kitaev terms in the undistorted case, hence no magnon gap at $K$ is present in the bare magnon dispersions shown in panels (a) and (c) of Fig.~\ref{fig:polaritons-theo}, in contrast to TDDFpT magnon dispersions that show a gap of 0.47~meV (see Fig.~\ref{fig:TDDFpT_spectrum}). While the inclusion of these SOC effects in the model Hamiltonian of the undistorted lattice is of great interest and possibly the topic of future studies, our main goal here is to investigate the sheer effect of the spin-lattice coupling on the gap opening around $K$.
As can be seen in Figs.~\ref{fig:polaritons-theo}(b) and (d), the main polaritonic effect is due to the $13$-th to $16$-th phonon modes, which open a gap of roughly $\approx 1.5$~meV, but
less intense hybridization features exist at lower energy as well.
The net outcome is the suppression of the magnon character in an energy region of about $4$~meV, which implies a concomitant suppression of the intensity in INS experiments. We remark that the exact location of this region in the BZ depends on the spin stiffness, which density functionals presently available are not able to predict with sufficient accuracy~\cite{Ke:2021}. However, a renormalization of such stiffness to experimental data brings our theoretical prediction of both the energy and the location in the BZ at the $K$ point of the gap in closer agreement with experiments for CrI$_3$ thin crystals, as illustrated in Figs.~\ref{fig:polaritons-theo}(c-d), and in the three-dimensional plot of Fig.~\ref{fig:3D}.
%

\section{Conclusions}
\label{sec:Conclusions}

The work reported in this paper was originally motivated by our effort to explain the gap observed in the spin-wave spectra of CrI$_3$ multi-layer samples~\cite{Chen:2018,Chen:2021}, which was overlooked in our first attempt to predict it from TDDFpT calculations performed for a mono-layer at clamped nuclei, as reported in the first version of this paper~\cite{Delugas:2021}. 
This failure was later found to be due to slight numerical inaccuracies in the computer code used to perform the computations, which were fixed since. 
The fix resulted in the opening of a gap at the $K$ point of the mono-layer BZ, which was however way too small with respect to experimental findings. A proper consideration of intra-layer interactions, implicitly accounted for in our subsequent computations performed for the bulk~\cite{Gorni:2023}, considerably enhances the value of the gap, still failing to match the experimental value. While the remaining inaccuracy may call for different explanations, starting from the inadequacy of the LSDA energy functional being employed, the results reported in the present paper for the mono-layer show that spin-lattice interactions may be at play in the bulk as well and decisively contribute to the opening of the observed gap. Whether or not this is the case certainly deserves further theoretical and experimental investigations. 
On the theoretical side, it will be interesting to quantify as well the impact of higher-order spin-lattice couplings, which may significantly affect the magnon dispersions and lifetimes~\cite{Cong:2022}.
On the experimental side, it would be important to have direct access to magnon dispersions in the monolayer regime. Also, it would be nice to tune the gap in the bulk by varying the strength of interlayer couplings, by either intercalation of some inert atomic species, or by application of a uniaxial pressure, which would all represent precious experimental input for the understanding of magnetic interactions in 2D magnets.

\section*{Acknowledgements}

This work was partially funded by the European Union through the \emph{ \textsc{MaX} Centre of Excellence for Supercomputing applications} (project No. 824143), by the Italian MUR through the PRIN 2017 \emph{FERMAT} (grant No. 2017KFY7XF) and the \emph{National Centre for HPC, Big Data, and Quantum Computing} (grant No. CN00000013), and by the Swiss National Science Foundation (SNSF), through grant No. 200021-179138, and its National Centre of Competence in Research (NCCR) MARVEL.

\appendix

\section{Computational details}
\label{App:computational}

All the calculations were performed using the \qe\ distribution~\cite{Giannozzi:2009, Giannozzi:2017, Giannozzi:2020}, a plane-wave+pseudopotential suite of computer codes, using the LSDA exchange-correlation functional and including SOC self-consistently by means of fully-relativistic pseudopotentials (FRPPs). We have used the norm-conserving FRPPs from the PseudoDojo library~\cite{vanSetten:2018} and by generating FRPPs with the \texttt{atomic} code using the configurations from v0.3.1 of the PSlibrary~\cite{DALCORSO2014337}. 

For the ground-state calculations, the Kohn-Sham wavefunctions and potentials were expanded in plane waves up to a kinetic-energy cutoff of 80\,Ry and 320\,Ry, respectively. Brillouin zone was sampled using a uniform  $\Gamma$-centered $8\times 8\times 1$ $\bm{k}$ points mesh for the hexagonal unit cell; uniform meshes of the same density have been adopted for calculations with supercells. The CrI$_3$ 2D crystal structure was obtained by extracting one layer from the trigonal bulk structure and by optimizing atomic positions and the in-plane lattice constant. The optimized in-plane lattice constant is 12.98~Bohr. The DFT calculations correctly yield a ferromagnetic (FM) ordering and a magnetic anisotropy with the out-of-plane directions as the easy axis.

Phonon dispersions were obtained using DFpT by computing the dynamical matrices on a $\Gamma$-centered $4\times 4 \times 1$ $\bm{q}$~points mesh; these results were then  used to compute the matrix of interatomic force constants (IFC) in real space from which phonon energies and displacements at arbitrary wavevectors were derived. The electrostatic long-range part of the IFC was computed by taking into account the artifacts produced by the nonphysical periodicity in the out-of-plane direction~\cite{sohier17}. 

All the calculations of the magnon dispersions have been performed using the \texttt{turboMagnon} component~\cite{Gorni:2022b} of the \qe\ suite of codes~\cite{Giannozzi:2009,*Giannozzi:2017,*Giannozzi:2020}, implementing the Liouville-Lanczos approach to TDDFpT in the adiabatic LSDA and including SOC self-consistently. This approach does not require the computation of unoccupied KS states and allows to evaluate the spin susceptibility matrix in Eq.~\eqref{eq:S_def}, ${\boldsymbol \chi}(\bm{q},\bm{q}; \omega)$, using three Lanczos chains per wavenumber. 
With respect to what is detailed in Ref.~\cite{Gorni:2018}, the algorithm has been upgraded by implementing the pseudo-hermitian symmetry along the lines of Ref.~\cite{Gruning:2011}, yielding converged spectra with $\approx 20000$ Lanczos steps, each step having roughly the cost of two Hamiltonian builds in a conventional static DFpT calculation. A Lorentzian smearing function with a broadening parameter of $0.03$\,meV has been used in the post-processing calculation. For the TDDFpT calculations, the kinetic energy cutoff is set to 60 and 240\,Ry for the wavefunctions and potentials, respectively. The same $\bm{k}$ points mesh as for ground-state calculations has been used.

The data used to produce the results of this work are available in the Materials Cloud Archive~\cite{MaterialsCloudArchive2023}.

\subsection{Exchange interactions}
\label{App:Jexch}

\begin{figure}[t]
    \centering
    \includegraphics[width=0.25\textwidth]{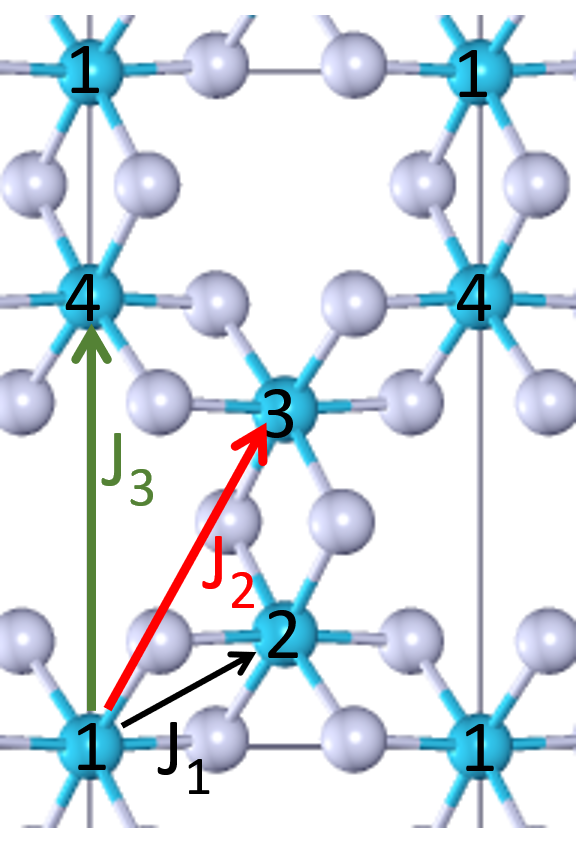}
    \caption{Supercell used for the estimate of exchange coupling parameters. The 4 inequivalent Cr sites (in blue) are numbered from 1 to 4 for further reference. The arrows indicate the couplings between nearest neighbors ($J_1$), next nearest neighbors ($J_2$), and third nearest neighbors ($J_3$).}
    \label{fig:cella_j}
\end{figure}

The estimate of the exchange interaction parameters between the Cr magnetic moments has been done by computing total energy differences between the FM ground state and three antiferromagnetic (AF) patterns contained in the supercell shown Fig.~\ref{fig:cella_j}. The details of these four configurations are reported in Table~\ref{tab:patterns}.

\begin{table}[h]
    \centering\
    \begin{tabular}{l c c c c c c}
    \hline
    \hline
    Pattern & $S_1$ & $S_2$ & $S_3$ & $S_4$&E$_{cell}^\mathrm{SOC}$ (meV)  & E$_{cell}^\mathrm{S.R.}$ (meV) \\
    \hline
    FM   & 3/2     & 3/2     & 3/2     & 3/2    &  0.00   &     0.00\\
    AFM  & 3/2     & -3/2    & 3/2     & -3/2   &  67.08  &    70.89\\ 
    AFX  & 3/2     & -3/2    & -3/2   & 3/2    &   78.94  &    82.80\\
    AF2Y & 3/2     & 3/2    & -3/2    & -3/2   &   46.82  &    49.17\\
    \hline
    \hline
    \end{tabular}
    \caption{The four magnetic arrangements used to estimate the exchange couplings. $\{S_i\}$ are the values of the local spins for each arrangement. The ``SOC'' suffix indicates  the energies per cell obtained taking into account SOC, and the ``S.R.'' suffix indicates instead the energies per cell obtained with the scalar relativistic approximation discarding SOC. All energies are reported relatively to the FM ground state. AFM, AFX, and AF2Y indicate three different AF arrangements.}
    \label{tab:patterns}
\end{table}

In order to estimate the exchange parameters from total energy differences one needs to express the total energy of each magnetic configuration as a function of the local moments. We start from the Heisenberg Hamiltonian, Eq.~\eqref{eq:spin-ham}, and  write the total energy per cell as:  
\begin{equation}
  E_{tot} = E_0 - \frac{1}{2} \sum^{cell}_{i} \sum_{j \ne i}^{all} J_{ij} S_i S_j ,
  \label{eq:toten0}
\end{equation}
where $E_0$ is an adjustable energy term, independent from  the spin interaction,  $J_{ij}$ are the isotropic exchange interaction parameters between spins $S_i$ and $S_j$ of sites $i$ and $j$, respectively, $\sum^{cell}$ indicates a sum over the spins contained in one unit cell, and $\sum^{all}$ may in principle range infinitely over the lattice. Assuming that the exchange interactions vanish over the third neighbours, we replace the coupling constants $J_{ij}$ with $J_1$, $J_2$, and $J_3$ for the couplings between first, second, and third nearest neighbours, respectively. 

\begin{table}[h]
    \centering
    \begin{tabular}{c c c c}
        \hline
        \hline
          Site &  \parbox{2.2cm}{$J_1$ (N.N.)} & \parbox{2.2cm}{$J_2$ (N.N.N)}  & \parbox{2.2cm}{$J_3$ ($3^\mathrm{rd}$ N.N)} \\ \hline
           1   &   2,2,4       &  3,3,3,3,1,1   &   4,4,4 \\ 
           2   &   1,1,3       &  4,4,4,4,2,2   &   3,3,3 \\
           3   &   4,4,2       &  1,1,1,1,3,3   &   2,2,2 \\
           4   &   3,3,1       &  2,2,2,2,4,4   &   1,1,1 \\ 
     \hline
     \hline
    \end{tabular}
    \caption{List of the nearest neighbors up to the third shell of each Cr site. Each site is indicated by the corresponding numeric label (see Fig.~\ref{fig:cella_j}).} 
    \label{tab:couplings}
\end{table}

The free parameters in Eq.~\eqref{eq:toten0} now reduce to four: $E_0, J_1, J_2, J_3$ to be fitted with the energy differences. Using the neighbours' list in Table~\ref{tab:couplings} and the spin and energy values from Table~\ref{tab:patterns}, one can write the following linear system: 
\begin{equation}
\begin{array}{cccccl}
   2E_0&-27J_1&-54J_2&-27J_3&=&2E_\mathrm{FM}, \\
   2E_0&+27J_1&-54J_2&+27J_3&=&2E_\mathrm{AFY}, \\
   2E_0&+9J_1&+18J_2&-27J_3&=&2E_\mathrm{AFX}, \\ 
   2E_0&-9J_1&+18J_2&+27J_3&=&2E_\mathrm{AF2Y}. 
\end{array}\
\label{eq:jeq}
\end{equation}
The resulting parameters, reported in Table~\ref{tab:j_values}, are in close agreement with those obtained by, Besbes et~al.~\cite{Besbes:2019} who used the magnetic force theorem. 
Together with  the value obtained for the onsite magnetic anisotropy $D^{zz}_s$ of 0.30~meV, the exchange parameters derived without SOC reproduce rather closely the fully-relativistic TDDFpT magnon dispersion of Fig.~\ref{fig:TDDFpT_spectrum}, with the exception of the gap at the Dirac point due to the lack of DM or Kitaev terms in our model.
Anisotropic exchange interactions have indeed been found to be nearly vanishing in the undistorted CrI$_3$ monolayer~\cite{Kvashnin:2020,Pizzochero:2020,Olsen:2021}, in agreement with our TDDFpT simulations, and their irrelevance in the  centro-symmetric system is also confirmed by the very small energy differences that we obtain inverting nonsymmetric magnetization patterns in the centro-symmetric structure. Moreover, at variance with the distorted structures, the centrosymmetric system always converges to ferromagnetic configurations with all spin parallel. In order to obtain  magnetization patterns that are not trivially equivalent to their inverted pattern we have to use constrained DFT  forcing the  magnetization to the selected patterns. 
Using this technique we have verified that for magnetic patterns as the one  depicted in Fig.~\ref{fig:mode-16-comparison}(c) --made of an equal weight superposition of a mode with  wave vector $\bm{q}_K$ on one sublattice and  one with wavevector $-\bm{q}_K$ on the other-- the constrained energies remain unchanged when the magnetic configuration is inverted either by switching sublattices or by inverting the sign of the amplitude $\theta$. 

\begin{table}[t!]
    \centering
    \begin{tabular}{lccc}
    \hline
    \hline
    Functional             & \parbox{1.5cm}{$J_1$} & \parbox{1.5cm}{$J_2$} & \parbox{1.5cm}{$J_3$} \\ \hline
    LSDA (no SOC)          &  $2.76$               & $0.81$                & $-0.27$  \\
    LSDA (with SOC)        &  $2.90$               & $0.85$                & $-0.28$  \\
    GGA~\cite{Besbes:2019} &  $2.37$               & $0.84$                & $-0.26$  \\
    \hline
    \hline
    \end{tabular}
    \caption{Isotropic exchange couplings between the Cr magnetic moments (in meV), according to the notation used in Eq.~\eqref{eq:spin-ham}. The absolute value of the Cr moments assumed in the model corresponds to a $S=3/2$ spin state.}
    \label{tab:j_values}
\end{table}
%

\subsection{Spin-lattice coupling constants}
\label{App:spin-lattice-constants}

The magnon-phonon coupling coefficient $\lambda_{\mu\nu}(\bm{q})$ have been computed with a sequence of ground-state, supercell calculations, by distorting the lattice along a given phonon mode $\mu\bm{q}$:
\begin{equation}
\bm{u}_{ns}
=
\xi^{\mu}(\bm{q})
\bigg[
\bm{e}^{\mu}_{s}(\bm{q}) e^{i\bm{q}\cdot\bm{R}_n}
+
\bm{e}^{\mu}_{s}(-\bm{q}) e^{-i\bm{q}\cdot\bm{R}_n}
\bigg]
\, ,
\label{eq:distortion}
\end{equation}
and then by projecting the magnetization pattern on the spin-wave normal modes at the lattice equilibrium to find the spin-wave amplitudes $\eta^{\nu}(\bm{q})$.
Considering a coupling in the form of Eq.~\eqref{eq:spin-lattice-ham}, the value of $\lambda_{\mu\nu}(\bm{q})$ can be found from the stationarity condition for the spin-wave amplitudes $\dot{\eta}^{\nu}(\bm{q}) = 0$ at a fixed phonon amplitude $\xi^{\mu}(\bm{q})$ (which can be considered to be a real number due to the presence of inversion symmetry in the undistorted system), yielding
\begin{equation}
    \lambda_{\mu\nu}(\bm{q}) = \sqrt{\frac{\omega_{\nu\bm{q}}}{2}} \frac{\eta^{\nu}(\bm{q})}{2\xi^{\mu}(\bm{q})} \, .
    \label{eq:lambda}
\end{equation}

\begin{figure*}[t]
    \begin{center}
        \includegraphics[width=0.98\textwidth]{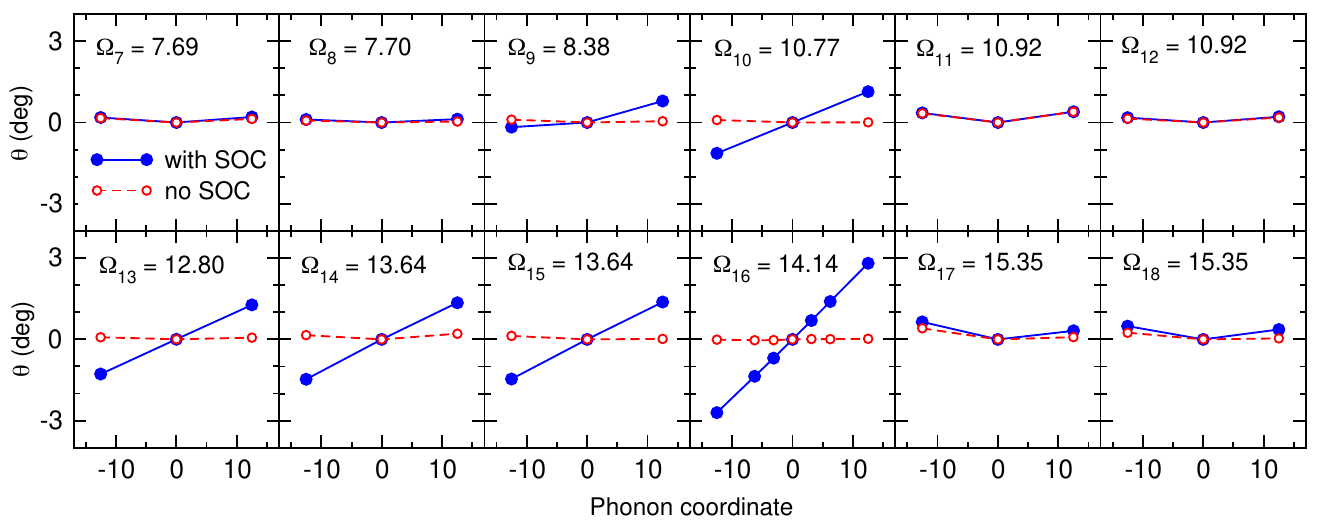}
        \caption{Average changes in the angle $\theta$ (in degrees) between the $z$ axis and the spin momentum on Cr atoms, which are induced due to the displacement of Cr atoms along the $\mu$-th normal phonon mode ($\mu=7,...,18$) with the amplitude $\xi^\mu(\bm{q}_K)$ [see Eq.~\eqref{eq:ph-normal-modes}], with and without SOC. On each panel, the phonon frequency of the corresponding $\mu$-th phonon mode at the Dirac point [$\Omega_\mu(\bm{q}_K)$ in meV] is indicated. The abscissa of all panels is the normalized phonon coordinate, $\xi^\mu(\bm{q}_K)$ [see Eq.~\eqref{eq:ph-normal-modes}], in units $\mathrm{\AA\times\sqrt{\mathrm{AMU}}}$.}
        \label{fig:Theta-of-Q}
    \end{center}
\end{figure*}

In order to identify a linear magnon-phonon coupling from these ground-state calculations, the value in Eq.~\eqref{eq:lambda} must not depend on the phonon amplitude $\xi^{\mu}(\bm{q})$.
When distorting along the 16-th phonon mode at the Dirac point, we find the magnetization to be a linear superposition of the two degenerate magnon modes with wave vector $\pm \bm{q}_K$, consistently with the fact that the lattice distortion includes Fourier components of $\pm \bm{q}_K$, as shown in Eq.~\eqref{eq:distortion}.
The two magnon branches are found to have equal weight, and to be in phase at $\bm{q}_K$ and in counterphase at $-\bm{q}_K$, leading the $+\bm{q}_K$-component to couple with the Cr$_1$ sublattice, and the $-\bm{q}_K$ component to couple with the Cr$_2$ sublattice , as shown in Fig.~\ref{fig:mode-16-comparison}(c). Formally one can write
\begin{multline}
    S^+_{ns} = \eta 
    \Bigg[ \bigg ( f^{\rm acu}_{s}(\bm{q}_K) + f^{\rm opt}_{s}(\bm{q}_K) \bigg) e^{i\bm{q}_K\cdot\bm{R}_n} \\
    + e^{i \phi} \bigg( f^{\rm acu}_{s}(-\bm{q}_K) - f^{\rm opt}_{s}(-\bm{q}_K) \bigg) e^{-i \bm{q}_K\cdot\bm{R}_n} \Bigg] \, , \label{eq:mag-decomp}
\end{multline}
where the phase difference between the modes at $\pm \bm{q}_K$ is found to be $\phi\approx-80^{\circ}$ and the absolute value of $\eta$ is related to the polar angle of the sublattice magnetization via $\sin(\theta_{ns}) = |S^+_{ns}|/S = |\eta|/\sqrt{2S}$.
For a given distorsion amplitude, all the magnetic moments inside the supercell are found to show the same deviation $\theta_{ns}$ within a $0.5^{\circ}$ precision, consistently with our assumption of a linear coupling. In the following, only $\theta$, the supercell average of $\theta_{ns}$, will therefore be reported.
We show the computed dependence of the spin polar angle with respect to the 16-th phonon amplitude in Fig.~\ref{fig:mode-16-comparison}(d).
The linear dependence for small angles, together with the lack of magnetization response without SOC, confirms the linear character between the spin and lattice degrees of freedom, the slope yielding the coupling magnitude $|\eta|$ according to Eq.~\eqref{eq:lambda}.

We performed similar calculations for all the phonon branches between $6$ and $25$~meV, whose $\theta_{ns} = \theta_{ns}(\xi^\mu(\bm{q}_K))$ dependences are reported in Fig.~\ref{fig:Theta-of-Q}.
All the phonon modes that couple linearly with the magnetization are found to induce a magnetization response in the form of Eq.~\eqref{eq:mag-decomp}.
The resulting magnitudes of the magnon-phonon coefficients $|\hbar\lambda_{\mu\nu}(\bm{q}\!=\!\bm{q}_K)|$ are reported in Fig.~\ref{fig:couplings}, showing a stark increase in the coupling intensity roughly in correspondence of the increase in the phonon vDOS, pointing towards a relevant effect of the magnon-phonon coupling in proximity of the Dirac point.

\section{Derivation of the spin-lattice Hamiltonian}
\label{App:Hamiltonian_mp}
In this appendix we present a derivation of the coupling term of Eq.~\eqref{eq:spin-lattice-ham} in our spin-lattice Hamiltonian.
Starting from Eq.~\eqref{eq:spin-ham}
\begin{equation}
    H_{\rm sp} = -\frac{1}{2}\sideset{}{'}\sum_{i \neq j} \bm{S}_{i} \cdot \bm{\mathcal{J}}_{ij} \cdot \bm{S}_{j} \\ -\sideset{}{'}\sum_{i} \bm{S}_{i} \cdot \bm{\mathcal{D}}_{i} \cdot \bm{S}_{i} \, ,
\end{equation}
where $i\equiv ns$ and $j\equiv mt$, we perform an expansion around the ferromagnetic state with the magnetization polarized along $z$:
\begin{align}
S_i^z
\approx
S - \frac{1}{2S}S^+_iS^-_i
\,.
\end{align}
Recalling that $S_i^{\pm} = S^x_i \pm i S^y_i$, one obtains
\begin{align}
H_{\rm sp}
=
&
\:
E_{\rm FM}
-
S \sideset{}{'}\sum_i \Big( a_i S^+_i + {\rm c.c.} \Big)
\nonumber\\
&
-
\sideset{}{'}\sum_i \Big( b_i S^+_i S^+_i + {\rm c.c.} \Big)
-
\sideset{}{'}\sum_i c_i S^+_i S^-_i
\nonumber\\
&
- \frac{1}{2}\sideset{}{'}\sum_{ij} \Big( d_{ij} S^+_i S^+_j + {\rm c.c.} \Big)
\nonumber\\
&
- \frac{1}{2}\sideset{}{'}\sum_{ij} \Big( e_{ij} S^+_i S^-_j + {\rm c.c.} \Big)
+\mathcal{O}\bigg[ \Big(S^+_iS^-_j\Big)^2\bigg]
\, ,
\label{eq:small-osc-Ham}
\end{align}
with $E_{\rm FM} = -S^2 \sum'_i \Big( \mathcal{D}_i^{zz} + \frac{1}{2}\sum'_j {\mathcal{J}}^{zz}_{ij}\Big)$ being the energy of the ferromagnetic state and
\begin{align}
a_i &= \mathcal{D}_i^{xz} - i \mathcal{D}_i^{yz} + \frac{1}{2}\sideset{}{'}\sum_j \bigg( \mathcal{J}_{ij}^{xz} -i \mathcal{J}_{ij}^{yz}\bigg),
\\
b_i &= \frac{1}{4}\bigg( \mathcal{D}_i^{xx} -\mathcal{D}_i^{yy} \bigg) -i \frac{\mathcal{D}^{xy}_i}{2},
\\
c_i &=  \frac{1}{2}\bigg( \mathcal{D}_i^{xx} +\mathcal{D}_i^{yy} \bigg) - \mathcal{D}^{zz}_i -\frac{1}{2}\sideset{}{'}\sum_j \mathcal{J}^{zz}_{ij},
\\
d_{ij} &= \frac{1}{4}\bigg( \mathcal{J}_{ij}^{xx} -\mathcal{J}_{ij}^{yy} \bigg) - \frac{i}{4}\bigg( \mathcal{J}_{ij}^{xy} +\mathcal{J}_{ji}^{xy} \bigg),
\\
e_{ij} &= \frac{1}{4}\bigg( \mathcal{J}_{ij}^{xx} +\mathcal{J}_{ij}^{yy} \bigg) + \frac{i}{4}\bigg( \mathcal{J}_{ij}^{xy} -\mathcal{J}_{ji}^{xy} \bigg)
\, .
\end{align}
From Eq.~\eqref{eq:small-osc-Ham} one can see that an anisotropy-induced linear coupling between the spin and lattice degrees of freedom emerges from variations of the $a_i$ coefficient with respect to the lattice displacements. Coming back to an explicit lattice notation
\begin{align}
a_{ns}
&\approx
 a_{ns}\Big|_{\rm eq.} + \sum_{mt}\frac{\partial a_{ns}}{\partial (\bm{R}_m+{\boldsymbol \tau}_{t})} \bigg|_{\rm eq.}\cdot \bm{u}_{mt}
 \nonumber\\
 &=
 \sum_{mt} \bm{a}'_{st}(\bm{R}_{mn})\cdot \bm{u}_{mt}
 \, ,
\end{align}
with ${\boldsymbol \tau_t}$ denoting the position of the $t$-th spin inside a unit cell.
In the second line we imposed translational invariance and assumed that $ a_{ns}\Big|_{\rm eq.} =0$. Substituting this last term back into Eq.~\eqref{eq:small-osc-Ham} one can write the linear spin-lattice coupling in the form of
\begin{equation}
H_{\rm sp} = - S \sideset{}{'}\sum_{ns} \sum_{mt} \Big( \bm{a}'_{st}(\bm{R}_{mn})\cdot \bm{u}_{mt} S^+_{ns} + {\rm c.c.} \Big) \,.
\label{eq:small-osc-Ham_linear}
\end{equation}
Finally, by expanding $\bm{u}_{mt}$ and $S^+_{ns}$ in Eq.~\eqref{eq:small-osc-Ham_linear} on their normal modes according to Eqs.~\eqref{eq:ph-normal-modes} and \eqref{eq:mag-normal-modes}, we obtain the expression for the spin-lattice coupling strength of Eq.~\eqref{eq:spin-lattice-ham}, with:
\begin{equation}
 \lambda_{\mu\nu}(\bm{q}) \sqrt{\omega_{\nu}(\bm{q})} \equiv
 S \sideset{}{'}\sum_s \sum_t
f_s^{\mu *}(\bm{q}) \bm{a}'_{st}(\bm{q}) \cdot \bm{e}_t^{\nu}(\bm{q}) 
\, .
\end{equation}

\bibliography{biblio}

\end{document}